\documentclass[12pt,preprint]{aastex}

\shorttitle{Eclipsing binaries in NGC 6362}
\shortauthors{Kaluzny et al.}

\begin{document}

\date{\today}

\title{The cluster ages experiment (CASE). VII. Analysis of two 
eclipsing binaries in the globular cluster NGC 6362{\LARGE$^\ast$}
      }
\author{J. Kaluzny$^1$\dag, I. B. Thompson$^2$, A. Dotter$^3$, M. Rozyczka$^1$, 
        A. Schwarzenberg-Czerny$^1$}
\affil{$^1$\small{Nicolaus Copernicus Astronomical Center, Bartycka 18, 00-716 Warsaw, 
      Poland}; \footnotesize{mnr@camk.edu.pl, alex@camk.edu.pl}}
\affil{$^2$\small{Observatories of the Carnegie Institution of Washington, 813 Santa 
       Barbara Street, Pasadena, CA 91101-1292, USA}; 
       \footnotesize{ian@obs.carnegiescience.edu}}
\affil{$^3$\small{Research School of Astronomy and Astrophysics, 
Australian National University
      Canberra, Australia}; \footnotesize{aaron.dotter@gmail.com}}

\begin{abstract}
We use photometric and spectroscopic observations of the detached eclipsing
binaries V40 and V41 in the globular cluster NGC~6362 to derive masses,
radii, and luminosities of the component stars. The orbital periods of
these systems are 5.30 and  17.89 d, respectively. The measured masses of
the primary and secondary components ($M_p$, $M_s$) are (0.8337$\pm$0.0063,
0.7947$\pm$0.0048) M$_\odot$ for V40 and (0.8215$\pm$0.0058,
0.7280$\pm$0.0047) M$_\odot$ for V41. The measured radii ($R_p$, $R_s$) are
(1.3253$\pm$0.0075, 0.997$\pm$0.013) R$_\odot$ for V40 and
(1.0739$\pm$0.0048, 0.7307$\pm$0.0046) R$_\odot$ for V41. Based on the
derived luminosities, we find that the distance modulus of the cluster is
14.74$\pm$0.04 mag -- in good agreement with 14.72 mag obtained from CMD
fitting. We compare the absolute parameters of component stars with
theoretical isochrones in mass-radius and mass-luminosity diagrams. For
assumed abundances [Fe/H] = -1.07, [$\alpha$/Fe] = 0.4, and Y = 0.25 we find the
most probable age of V40 to be 11.7$\pm$0.2 Gyr, compatible with the age of
the cluster derived from CMD fitting (12.5$\pm$0.5 Gyr). V41 seems to be
markedly younger than V40. If independently confirmed, this result
will suggest that V41 belongs to the younger of the two stellar
populations recently discovered in NGC 6362. The orbits of both systems are
eccentric. Given the orbital period and age of V40, its orbit should have
been tidally circularized some $\sim$7~Gyr ago. The observed eccentricity is
most likely the result of a relatively recent close stellar encounter.


{\def\thefootnote{}\footnotetext{\dag Deceased}}

\let\thefootnote\relax\footnotetext
{$^{\mathrm{\ast}}$This paper includes data gathered with the 6.5-m Magellan Baade 
and Clay Telescopes, and the 2.5-m du Pont Telescope located at Las Campanas Observatory, 
Chile.}

\end{abstract}


\section{Introduction}
\label{sect:intro}

The Cluster AgeS Experiment (CASE) is a long term project conducted mainly
at Las Campanas Observatory with a principal aim of the detection and detailed
study of detached eclipsing binaries (DEBs) in nearby globular clusters
(GCs). The main goal is to measure masses, luminosities and radii of DEB
components with a precision of better than one per cent in order to determine GC
ages and distances, and to test stellar evolution models \citep{jka05}. The
methods and assumptions we employ 
follow the ideas of \citet{pac97} and \citet{tho01}. Thus far, we have
presented results for seven DEBs from four clusters: $\omega$ Cen, 47 Tuc,
M4 and M55. These were the first and, to the best of our knowledge, the
only direct measurements of the fundamental parameters of main-sequence and
subgiant stars in GCs. More details on the project and references to
earlier papers can be found in \citet{jka14a}.

The present paper is devoted to an analysis of two DEBs, V40 and V41,
belonging to the globular cluster NGC 6362. Radial velocity and photometric
observations are used to derive masses, luminosities, and radii of the
component stars. In Section \ref{sect:photobs} we describe the photometric
observations and determine orbital ephemerides. Section \ref{sect:spectobs}
presents spectroscopic observations and radial-velocity measurements. The
combined photometric and spectroscopic solutions for the orbital elements
and  the resulting component parameters are obtained in Section
\ref{sect:lca_sp}. In Section \ref{sect:age}, we compare the derived
parameters to Dartmouth stellar evolution models, with an emphasis on
estimating the age of the binaries. Finally, in Section~\ref{sect:discus}
we summarize our findings.

\section{Photometric observations}
\label{sect:photobs}

Photometric observations analyzed in this paper were
obtained at Las Campanas Observatory between 1995 and 2011 using the 2.5-m
du Pont telescope equipped with the TEK5 and SITE2K CCD cameras, and the
1.0-m Swope telescope equipped with the SITE3 CCD camera. The scale was 0.26
arcsec/pixel and 0.44 arcsec/pixel for the du Pont and Swope telescopes,
respectively. For each telescope the same set of $BV$ filters was used for
the entire duration of the project. 
The first likely eclipse events of V40 and V41 were observed in June 1991.
For the next eight years observations were conducted in a survey mode with 
the aim of detecting more eclipses and to establish ephemerides for both 
variables. The periodic variability of V40 and V41 was firmly established by
\citet{bat99}. 
Starting with the 2001 season, we attempted
to cover  predicted eclipses of V40 or V41. Several eclipses were observed
with both telescopes. The light curve analysis in Section \ref{sect:lca_sp}
is based exclusively on the du Pont data. The less accurate Swope
photometry was used solely to improve the ephemerides (the difference in
accuracy between the two telescopes resulted from the difference in the
light-gathering power and the poorer sampling of the  point spread function
of the Swope telescope). 

The core and half-light radii of NGC~6362 are $r_{c}=70${\arcsec} and
$r_{h}=123$\arcsec, respectively \citep{har96}. V40 is located in the
core region at a distance of $d=59${\arcsec} from the cluster center.
V41 resides at the half-light radius with a distance of 
$d=121$\arcsec. The du Pont images show no evidence of unresolved visual 
companions to either binary. In the case of V40 this is confirmed by the 
examination of HST/ACS images obtained within the program
GO-10775 (PI Ata Sarajedini). Both systems
are proper-motion members of NGC 6362 \citep{zlo12, jka14b}. Coordinates and
finding charts for V40 and V41 can be found in \citet{bat99}.

The light curves presented here were measured with profile fitting
photometry extracted with the  Daophot/Allstar codes \citep{ste87,ste90},
and transformed to the standard Johnson system based on observations of
\citet{land92} standards. A more detailed description of the data and the
reduction procedures can be found in a paper presenting the photometry of
69 variables from the field of NGC 6362 \citep{jka14b}.  We note that, in
general, more accurate photometry of crowded fields can be obtained with an
image subtraction technique. However, profile fitting  photometry is less
vulnerable to systematic errors, and as such it is a better option whenever
the star is free from blending. 

\subsection{ Ephemerides}
Several eclipse events of V40 were covered densely enough to determine 
individual times of minima. They were found using the KWALEX code which 
is based on an improved version of the algorithm presented by \citet{kwee}  
(see Appendix). The linear ephemerides:
\begin{eqnarray}
 HJD^\mathrm{Min I}& = &245 3915.50966(25) + 5.2961725(14)\times E\nonumber\\
 HJD^\mathrm{Min II}& = &245 3891.82853(14) + 5.29617486(99)\times E\label{eq:ephe40}
\end{eqnarray}
proved to be an adequate fit to the moments of minima listed in Table~\ref{V40_min}. 
The orbit of V40 is eccentric. Since the secondary 
(shallower) minimum was observed more times than the primary one, we decided to choose 
it as a base of the phase count. Thus, we set $\phi^\mathrm{Min II}=0.5$, which 
necessarily results in $\phi^\mathrm{Min I}\ne0$.

\begin{figure}[!h]
\centering
\includegraphics[width=0.90\textwidth,bb= 39 175 563 690,clip]{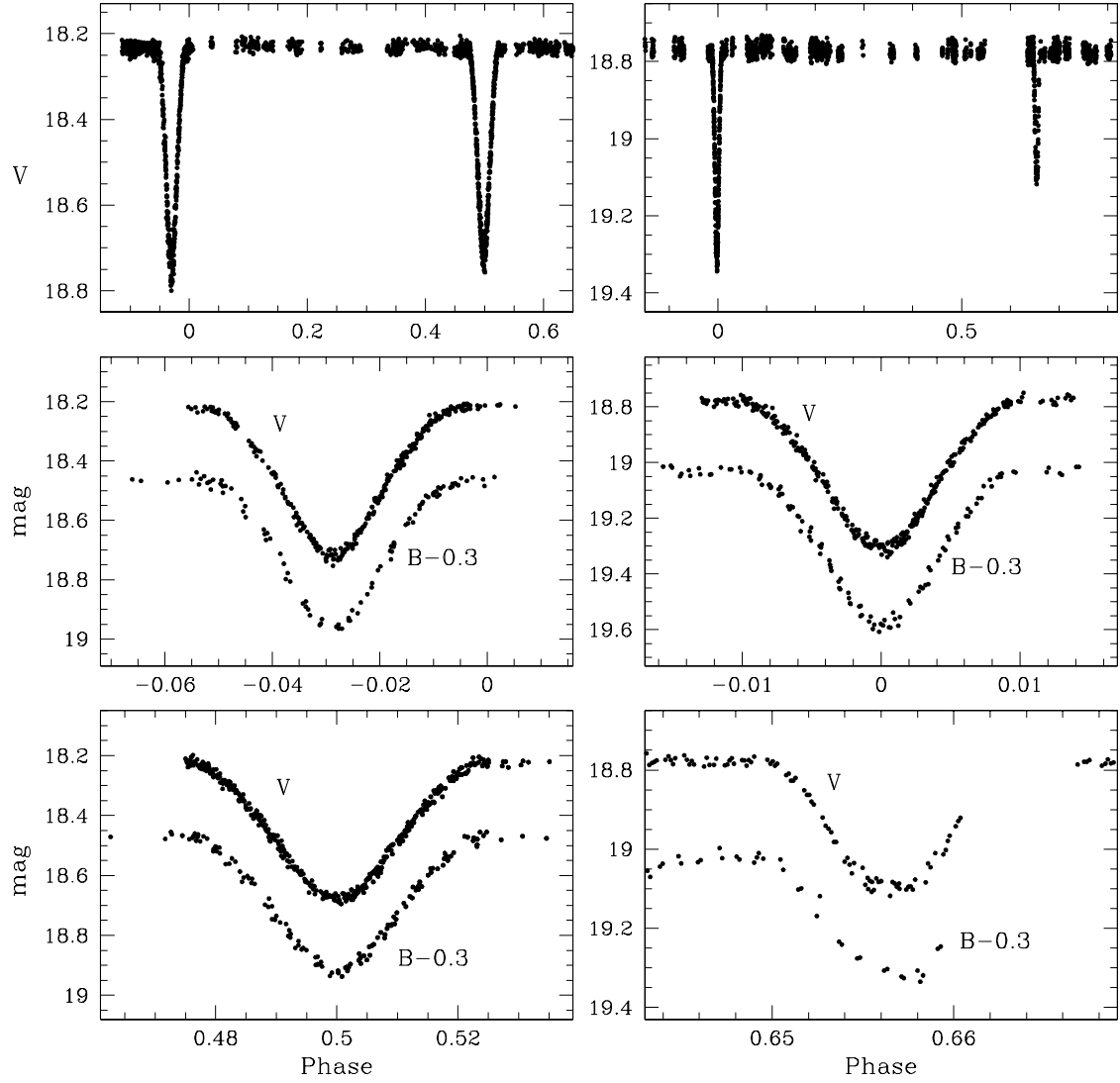}
\caption {Light curves adopted for the analysis: V40 (left column) and V41 
          (right column). In each column the complete curve is shown in the 
           top panel, followed by zoomed primary minimum and secondary minimum.
           The $B$-curves are shifted by 0.3 mag for clarity.
         }
\label{fig:lc}
\end{figure}

Altogether, 13 eclipses of V41 were observed between August 2007 and
September 2011. However, ingress and egress were only covered for a few of
them. This prevented us from a classical period study based on the
determination of individual moments of minima and the O-C technique. To
find the ephemeris, a combined $V$ light curve including data from all
seasons and both telescopes was used. This was fitted with the help of the
JKTEBOP~v34 code\footnote{described in detail at and available from
http://www.astro.keele.ac.uk/jkt/codes/jktebop.html} \citep[][and references therein]{south13},
allowing only for variations of $P$ and moment of the primary eclipse 
$T_{0}$. Mode 8 of the code was used, which enabled a robust determination
of the errors of the two parameters with the help of  Monte Carlo simulations.
The following linear ephemeris 
\begin{equation}
 HJD^\mathrm{Min I} = 245 4263.65242(25) + 17.8888441(40)\times E 
 \label{eq:ephe41} 
\end{equation}
was obtained. The light curves phased with ephemerides
given by equations (\ref{eq:ephe40}) and (\ref{eq:ephe41}), and used for
the subsequent analysis, are shown in Fig.~\ref{fig:lc}.

\section{Spectroscopic observations}
\label{sect:spectobs}

The spectra were taken with the MIKE echelle spectrograph \citep{bern03} on
the Magellan Baade and Clay 6.5-m telescopes, using a 0.7\arcsec\ slit
which provided a resolution R~$\approx$40,000. A typical observation
consisted of two 1800-second exposures of the target, flanking an exposure
of a thorium-argon hollow-cathode lamp. The raw spectra were reduced with
the pipeline software written by Dan Kelson, following the approach
outlined in \citet{kel03}. For the further processing, the IRAF package
ECHELLE was used. The velocities were measured using  software based on the
TODCOR algorithm \citep{zuc94}, kindly made available by Guillermo Torres.
Synthetic echelle-resolution spectra from the library of \citet{coelho06} 
were used as velocity templates. These templates were interpolated to the
values of $\log g$ and $T_{eff}$ derived from the photometric solution (see
Section \ref{sect:lca_sp}) and Gaussian-smoothed to match the resolution of
the observed spectra. For the interpolation in chemical composition we
assumed ${\rm [Fe/H]} = -1.07$ and $[\alpha/{\rm Fe}]=0.4$, i.e. values
adopted for the cluster as a whole \citep{car09, dotter10}. These values
are not well determined (see the discussion in Sect. \ref{sect:lca_sp}),
however the results of the velocity measurements are insensitive to minor
changes in [Fe/H] or $[\alpha/{\rm Fe}]$.

Radial velocities were measured on the  wavelength intervals 4120-4320 \AA\
and 4350-4600 \AA, covering the region of the the best signal-to-noise
ratio while avoiding the H$\gamma$ line. The final velocities, obtained
by taking the mean of these two measurements, are given in Tables \ref{V40vel} a
nd \ref{V41vel}, which list the heliocentric Julian dates at mid-exposure, 
velocities of the primary and secondary components, and the orbital phases of 
the observations calculated according to the ephemerides given by Equations (1) and (2).
An estimate of the error in the velocity measurements is the {\it rms} of 
half of the differences between velocities obtained from the two wavelength 
intervals. The corresponding values are 0.63, 0.59, 0.23 and 0.52 km/s for the primary 
and secondary of V40 and the primary and secondary of
V41, respectively. They are consistent with the {\it rms} values of 
the residuals from orbital fits given in Table \ref{param}. 

\section{Analysis of light and velocity curves}
\label{sect:lca_sp}

We assume that reddening is uniform for NGC~6362. \citet{har96} lists
$E(B-V)=0.09$ as measured  by West and Zinn (1984) based on the Q39 index.
The map of total galactic extinction by Schlegel et al. (1998) predicts
$E(B-V)=0.075$ mag for the position of NGC 6362. This value agrees with an
estimate resulting from  isochrone fitting  applied to HST/ACS photometry
of the cluster by \citet{dotter10}. They obtained $E(6-8)=0.070$,
equivalent to $E(B-V)=0.071$. This value is also compatible with values
obtained by \citet{kov} (${E(B-V)}=0.073\pm 0.007$) and \citet{ol01}
(${E(B-V)}=0.08\pm 0.01$) from an analysis of RR Lyr stars in NGC~6362. We
adopt $E(B-V)=0.075$ and neglect the differential reddening, which according 
to \citep{bon13} is 0.025 mag on the average. \citep{bon13} comment that this 
differential reddening  may result from uncorrected zero-point
variations in the original photometry. 

We employed a quadratic formula for limb darkening, and calculated 
coefficients iteratively together with effective temperatures of the
components of both systems. The temperatures were derived from dereddened
$B-V$ colors using the empirical calibration of \citet{cas10}. The
coefficients were then interpolated from the tables of \citet{claret00}
with the help of the JKTLD code.\footnote{Written by John Southworth and
available at www.astro.keele.ac.uk/jkt/codes/jktld.html} The updated colors
in turn were computed from total $V$ and $B$ magnitudes at maximum light
and light ratios derived from the light curve solution. We note that a
slightly more recent calibration by Sousa et al. (2011) results in
temperatures higher by 8~K on average, however the agreement between 
calibrations is excellent over the relevant range of colors.

We checked that using a square root or logarithmic formula for the limb
darkening instead of a quadratic formula had a negligible impact on the
photometric solution. In particular, relative radii derived from the three 
two-parameter formulas differed by less than $\sim$$0.2\sigma$), whereas
using a linear limb darkening formula resulted in a slightly poorer fit, 
and caused the relative radii to differ by $\sim$$1\sigma$ from those 
obtained with the quadratic formula. We
conclude that, for the stars analyzed here, two-parameter formulate provide
a better description of the limb darkening than the linear law. 

Light and radial velocity curves of V40 and V41 were analyzed with the JKTEBOP~v34 
code, which allows for a simultaneous
solution of one light curve and two radial velocity curves (one for the
primary and one for the secondary component). The $V$-curve was solved for 
the following parameters: inclination $i$, sum of fractional radii $r_p$ 
and $r_s$, ratio of the radii $r_{s}/r_{p}$; surface brightness ratio $S$ 
(secondary to primary), luminosity ratio $L_{s}/L_{p}$, $e\cos(\omega)$ and
$e\sin(\omega)$ where $e$ is the orbital eccentricity and $\omega$ the
longitude of the periastron, amplitudes of radial velocity curves $K_{p}$
and $K_{s}$, and systemic velocity $\gamma$. Errors of 
stellar parameters were derived using a Monte Carlo approach 
implemented in mode 8 of the JKTLD code. We performed 10000 Monte Carlo 
simulations for each of V40 and V41. 

\begin{figure}[!h]
\centering
\includegraphics[width=0.90\textwidth,bb= 46 431 564 690,clip]{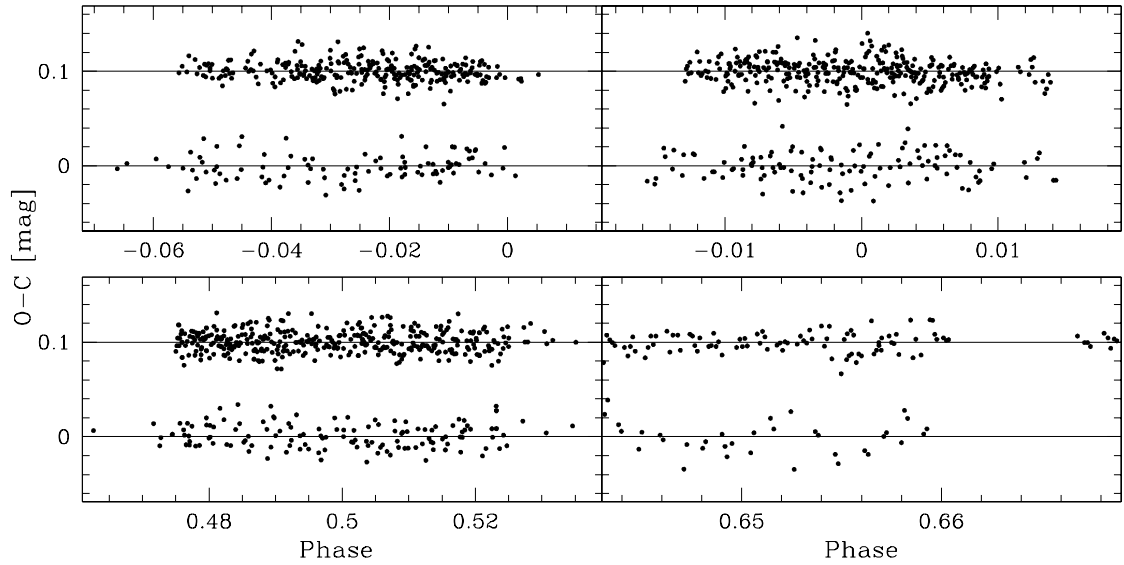}
\caption {Residuals from the fits to the light curves of V40 (left) 
          and V41 (right). The panels are arranged in the same way as the 
          two lower rows of Fig. \ref{fig:lc}. In each panel, the lower sequence 
          represents the $B$-residuals, and the upper one the $V$-residual
          offset by 0.1 mag for clarity.
         }
\label{fig:lcres}
\end{figure}

Since the $B$-data were of poorer quality than those collected in the $V$-band, 
we decided to use them solely to a measurement of 
the luminosity ratio $(L_s/L_p)_B$. This was accomplished by fixing all
parameters but the surface brightness ratio $S_B$. However, the formal error
of $S_B$ calculated by JKTEBOP would have been seriously
underestimated if uncertainties of other parameters had not been taken into
account. To avoid this problem we used the PHOEBE package v0.31a
\citep{prsa05}, which allows for a simultaneous solution of several light
curves but can be unstable when attempting to solve light and velocity
curves simultaneously. We initialized the package with the photometric 
solution found by JKTEBOP, loaded $BV$ curves, and, allowing all geometrical 
parameters to vary, found the standard error of $S_B$ using another Monte Carlo
procedure written in PHOEBE-scripter. The procedure replaces the
observed light curves $B_0$ and $V_0$ with the fitted values $B_f$ and
$V_f$ , generates Gaussian perturbations $\delta B_f$ and $\delta V_f$ such that 
the standard deviation of the perturbation is equal to the standard deviation of 
the corresponding residuals shown in Figure \ref{fig:lcres}, and performs PHOEBE 
iterations on $B_f + \delta B_f$ and $V_f + \delta V_f$. As before, 10000 Monte 
Carlo simulations were performed for each binary. In our opinion such a 
hybrid approach based on both the JKTEBOP and PHOEBE codes is optimal when
analyzing data sets similar to those presented in this paper.

\begin{figure}[!h]
\centering
\includegraphics[width=0.90\textwidth,bb= 43 420 563 690,clip]{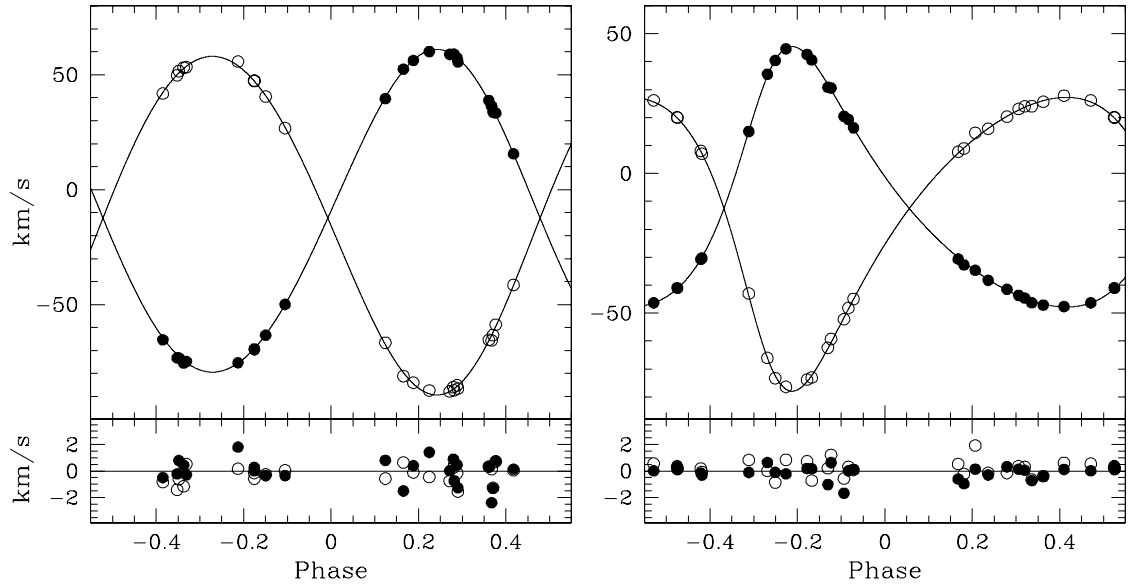}
\caption {Observed velocities of the components of V40 (left) and V41 (right) shown together 
          with the final fits. Filled and open symbols denote primary and secondary components, 
          respectively (note that for V40 phase 0.0 corresponds to the secondary minimum). 
          Errors of individual points are smaller than the symbols themselves.
          Lower panels display residuals from the fits. 
         }
\label{fig:vc}
\end{figure}

\begin{figure}[!h]
\centering
\includegraphics[width=0.55\textwidth,bb= 43 148 569 690,clip]{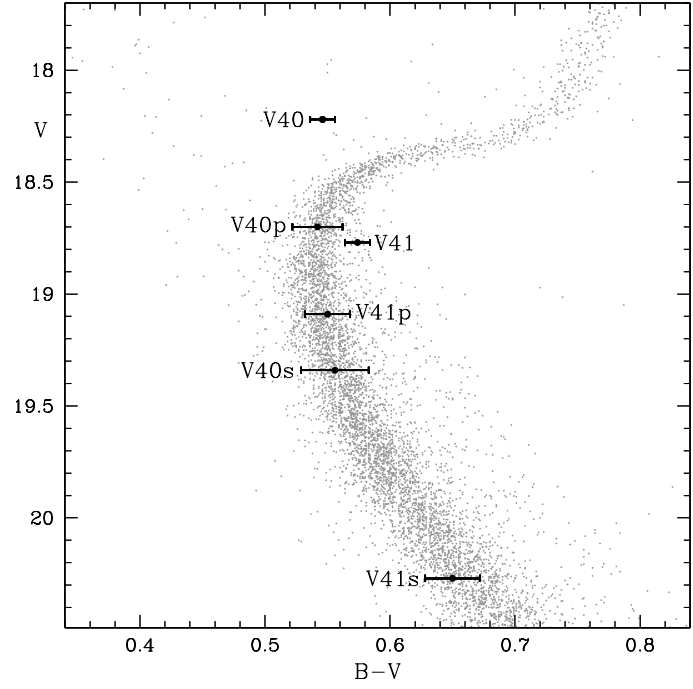}
\caption {The color-magnitude diagram of NGC~6362 with locations of V41, V42 and  
          their components. Data 
          for the background stars are taken from \citet{zlo12} (only proper motion  
          members of the cluster are shown).
         }
\label{fig:cmd}
\end{figure}

The quality of the final fits is illustrated in Figures \ref{fig:lcres} and
\ref{fig:vc}. Orbital and photometric parameters  derived for both systems
are presented in Table \ref{param}, and absolute parameters of the
components are listed in Table \ref{param2}. The errors of the effective
temperatures include uncertainties of observed colors, reddening, and
photometric zero points (0.015 mag for $V$ and $B-V$). The bolometric
luminosities are calculated from radii and temperatures adopting
$L_\odot^{bol}=3.827\times10^{33}$ ergs.\footnote{
https://sites.google.com/site/mamajeksstarnotes/basic-astronomical-data-for
-the-sun} To derive absolute visual magnitudes, we used bolometric
corrections based on atmosphere models of \citet{cas04} which ranged from
$-0.07$ mag for the primary of V40 to $-0.10$ mag for the secondary of V41.
The absolute visual magnitudes along with the $V$-magnitudes derived from
light curve solutions allowed for a determination of the apparent distance
moduli listed in Table \ref{param2}. The location of the components of V40
and V41 on the CMD of NGC 6362 is shown in Figure \ref{fig:cmd}. All four
stars are placed on the main sequence: three of them at the turnoff region,
and the fourth one (the secondary of V41) well below the turnoff.

\section{Age and distance analysis} 
\label{sect:age}

The fundamental parameters obtained from the analysis of V40 and V41 allow
us to derive the ages of the components of these systems, using theoretical
isochrones. However, before a model comparison is made it is necessary to
review the information available on the chemical composition of NGC~6362
since the model-based ages are sensitive to the adopted values of helium
abundance, [Fe/H] and [$\alpha$/Fe].

Unfortunately, the metallicity of NGC~6362 is not well known. A literature
search reveals that determinations of [Fe/H] from high resolution spectra
are limited to the study by \citet{car97}, who  found ${\rm [Fe/H]}=-0.96$
and ${\rm [Fe/H]}=-0.97$ for two red giants. \citet{car09} used this
determination along with some older estimates based on the integrated
photometry of NGC~6362 to derive ${\rm [Fe/H]}=-1.07\pm 0.05$ on their
newly proposed metallicity scale. 
For the sake of completeness, we note
that \citet{har96} listed ${\rm [Fe/H]}=-0.99$ -- a value obtained by
combining that of \citet{car09} with ${\rm [Fe/H]}=-0.74\pm 0.05$ derived
by Geisler et al. (1997) based on Washington photometry of 10 cluster
giants. \citet{ol01} obtained a helium content Y=0.292 from the analysis of
RRc stars. Finally, \citet{dal14} presented convincing evidence of multiple
populations in NGC~6362. We adopt a fiducial composition of [Fe/H]=$-1.07$,
[$\alpha$/Fe]=+0.4 and Y=0.25, but will investigate the dependence on these
parameters when deriving the ages of the component stars.

\begin{figure}[!h]
\centering
\includegraphics[width=\textwidth,bb= 34 324 563 689,clip]{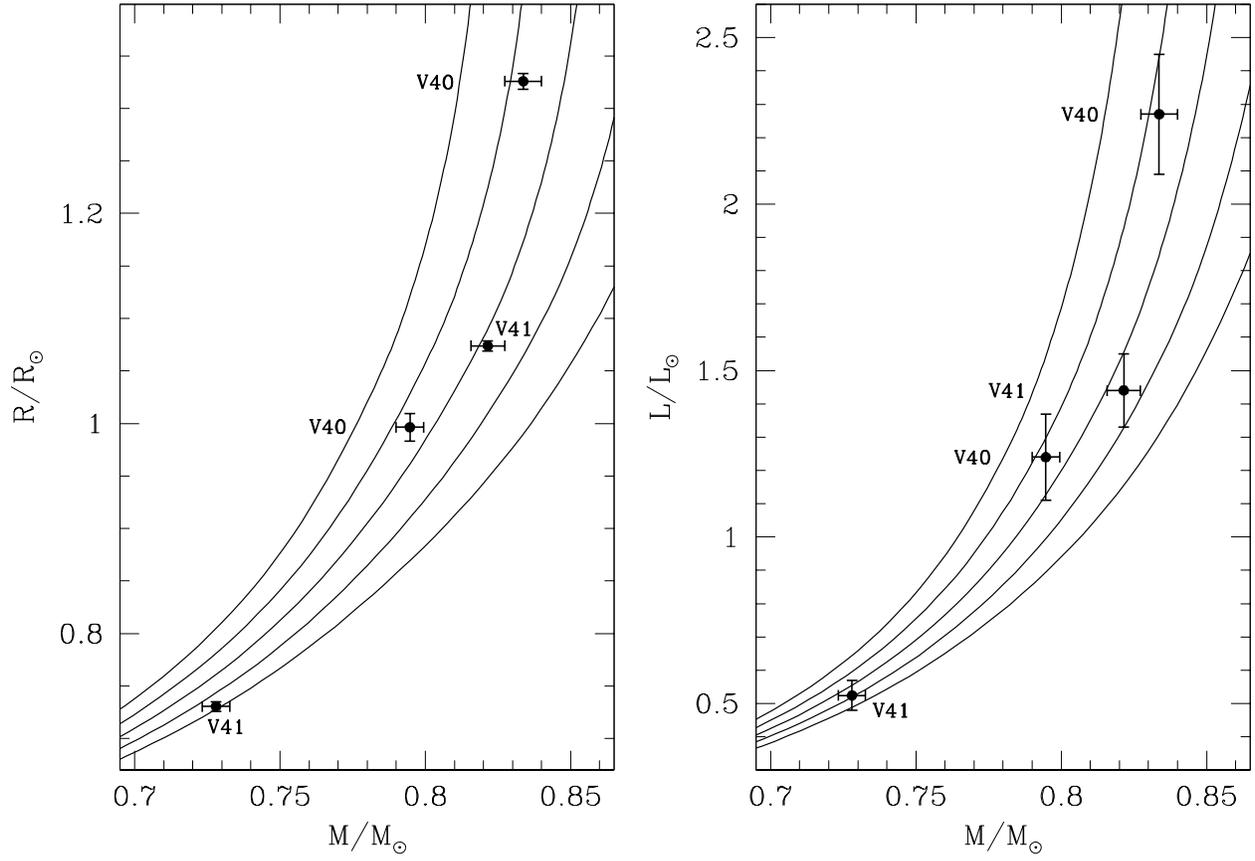}
\caption {Mass-radius and mass-luminosity diagrams for V40 and V41 components.
          Lines are DSED isochrones with Y=0.2496, [Fe/H]=-1.07 and 
         [$\alpha$/Fe]=0.4  for 9, 10, 11, 12 and 13 Gyr from right to left.
         }
\label{fig:rmlm}
\end{figure}

In a study of three binary systems in M4 \citep{jka13} we used two sets of
theoretical isochrones: Dartmouth \citep[][henceforth DSED]{dotter08} and
Victoria-Regina \citep{vandenberg12}. Since the differences in the derived
ages turned out to be smaller than the uncertainties imposed by the
observational errors, the present analysis is based on DSED isochrones
only. Figure \ref{fig:rmlm} shows DSED isochrones for ages ranging from
9-Gyr to 13-Gyr plotted on mass-radius and mass-luminosity diagrams, together
with values for V40 and V41. Both diagrams indicate that V41 may be
younger than V40, and the mass-radius diagram suggests that the secondary 
of V41 is younger than the primary (or, if both components are of the same age, 
the secondary is too small compared to the primary).  

\begin{figure}[!h]
\centering
\includegraphics[width=0.55\textwidth,bb= 19 159 563 689,clip]{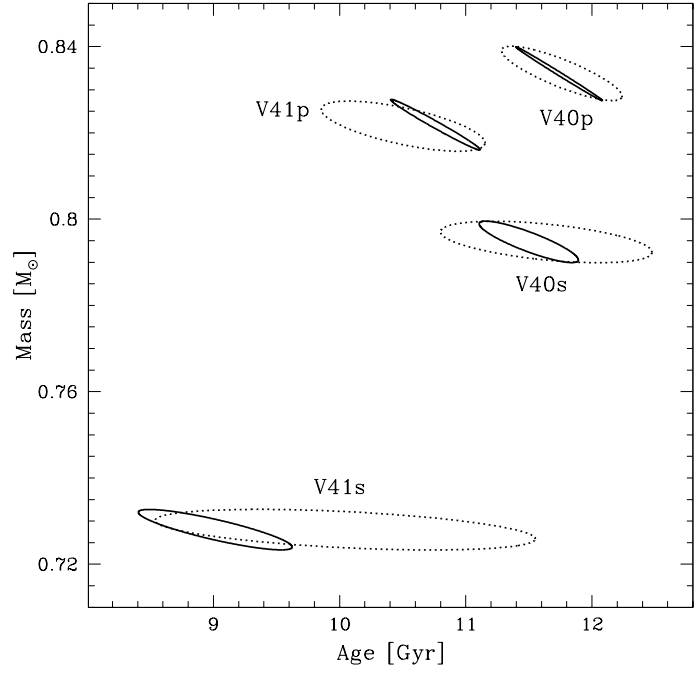}
\caption {Mass-age diagrams (MADs) for the components of V40 and V41. Results from the 
          analysis of $(M-R)$ and $(M-L)$ diagrams are shown with solid and dotted
          lines, respectively (see Section \ref{sect:age} for an explanation).
         }
\label{fig:ellipses}
\end{figure}

A detailed comparison of the ages of the individual components, which requires 
a quantitative age estimate from the data and the isochrones, was undertaken 
following  the approach developed
by \citet{jka13} to study the binary systems in M4. Briefly, the $(M-R)$ or
$(M-L)$ plane is densely sampled in both dimensions for each component
separately, and the age of an isochrone matching the properties of the
sampled point is found.  The age ensembles generated in this way are
visualized in the form of mass-age diagrams (MADs). The MAD of a given
component is composed of two ellipses derived from $(M-R)$ and $(M-L)$
relations. Each ellipse defines a range of ages achieved by stars whose
parameters differ from the corresponding entries $(M,R)$ or $(M,L)$ in
Table \ref{param2} by at most $\varepsilon_R$ or $\varepsilon_L$, where
$\varepsilon_R^2 = \sigma(M/M_\odot)^2 +\sigma(R/R_\odot)^2$ for the
$(M-R)$ plane and likewise for the $(M-L)$ plane; $\sigma(M/M_\odot)$,
$\sigma(R/R_\odot)$ and $\sigma(L/L_\odot)$ being the standard errors of
$M$, $R$ and $L$ from Table \ref{param2}. The MADs of the components of V40
and V41  are shown in Figure \ref{fig:ellipses}.

The age of each component was calculated as the weighted mean of 
the ages derived from the two corresponding MADs. For V40 we obtained 
11.7$\pm$0.3 Gyr for the primary and 11.5$\pm$0.4 Gyr for the secondary. The 
difference is 0.2$\pm$0.5 Gyr. For V41 the ages are 10.7$\pm$0.3 Gyr and 9.2$\pm$0.6 Gyr,
respectively, for the primary and the secondary. The difference is 1.5$\pm$0.7 Gyr.
Possible origins of this two-sigma effect are discussed in Section~\ref{sect:discus}. 
We estimated the age of each system from the weighted mean of the ages 
derived from each pair of corresponding MADs. For V40 we obtained 11.7$\pm$0.2 Gyr, 
in agreement with 12.5$\pm$0.5 Gyr derived by 
\citet{dotter10} from CMD fitting. For V41 we obtained 10.4$\pm$0.3 Gyr, which 
suggests that this system is significantly younger than V40, with a formal 
difference of 1.3$\pm$0.4 Gyr.

\begin{figure}[!h]
\centering
\includegraphics[width=0.55\textwidth,bb= 42 164 563 689,clip]{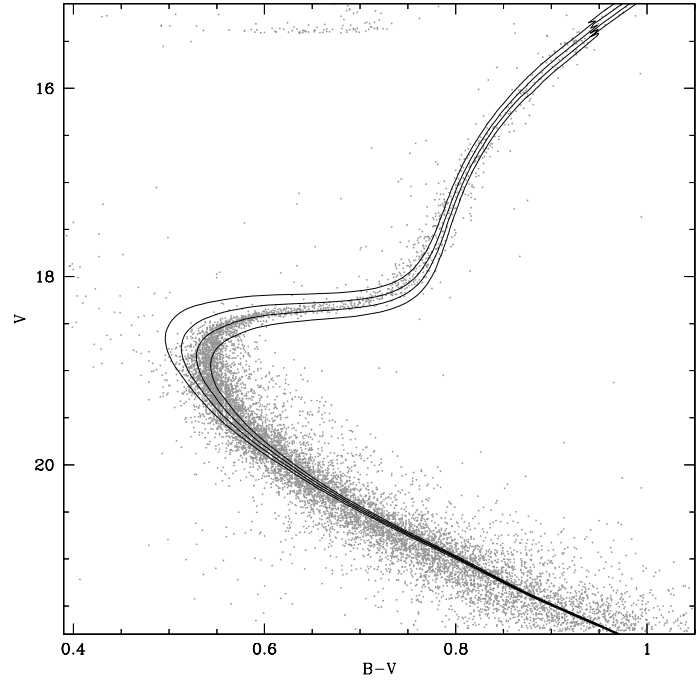}
\caption {CMD of NGC 6362 with DSED isochrones for 10, 11, 12 and 13 Gyr.
          Distance modulus and reddening resulting from the fit amount to 
          14.72 mag and 0.061 mag, respectively.
         }
\label{fig:cmd_iso}
\end{figure}

To independently estimate the age of the cluster as a whole, we fitted the CMD of NGC 
6362 with DSED isochrones (see Figure \ref{fig:cmd_iso}). Since uncertainties of stellar 
models grow larger with evolutionary time, when 
fitting we assigned the largest weight to main sequence and turnoff stars.
This results in an age estimate of 12.5$\pm$0.75 Gyr - again in agreement with the result of 
\citet{dotter10}, which is based on an independent set of photometric data. 
The CMD fit yielded a distance modulus of 14.72 mag - larger than 14.55 mag derived 
by \citet{dotter10}, but in good agreement with 14.68 mag listed by \citet{har96}. 
The reddening  derived from the CMD-fit ($E(B-V)=0.061$ mag) is slightly lower than the 
value of 0.075 mag adopted 
in Sect. \ref{sect:lca_sp}, but decreasing [Fe/H] from the fiducial -1.07 to -1.15 brings it back 
to the adopted value (we checked that such a small change in metallicity does not influence 
the parameters listed in Tables \ref{param} and \ref{param2} in any significant way). 
The distance moduli of all four stars forming V40 and V41 agree with each other (see Table \ref{param2}),
and the mean weighted modulus of 14.74$\pm$0.04 mag is consistent with that obtained from CMD fitting.

To sum up, the age derived from the MADs of V40 components is compatible with that 
derived from CMD fitting, however the age of the primary of V41 is $1.0\pm0.4$ Gyr 
younger than the age of V40, with the secondary of V41 being another $1.5\pm0.7$ Gyr younger. 
The problem looks much milder in the $(M-L)$ diagram, in which all the four components
are marginally coeval within 1-$\sigma$ limits. We stress, however, that the $(M-L)$ diagram 
and associated MADs must be given a smaller weight because the luminosities in Table \ref{param2} 
are sensitive to adopted chemical composition 
\citep[via $T_{eff}$, see][]{dotter10} and possible differential reddening, whereas the radii are entirely 
reddening and composition-independent.

\section{Discussion and summary} 
\label{sect:discus}

We derived absolute parameters of the components of V40 and V41, two detached eclipsing 
binaries located near the main sequence turnoff of NGC 6362. The accuracy of mass and radii 
measurements is generally better than 1 per cent, with the only exception being the radius of the 
secondary in V40 which is determined with an accuracy of 1.3 per cent. 
The luminosities of the components were found using effective temperatures estimated from 
$(B-V)$ - $T_\mathrm{eff}$ calibrations compiled by \citet{cas10}. 
While their errors are rather large (between 7 and 10 per cent), all the four distance moduli 
have nearly the same value, and the formally calculated mean modulus of 14.74$\pm$0.04 mag 
agrees very well with 14.72 mag derived from the comparison of DSED isochrones with the CMD 
of the cluster.  

Based on the parameters of V40 and DSED isochrones for Y=0.25, [Fe/H]=$-1.07$ and [$\alpha$/Fe]
=$+0.4$, we derived an age of 11.7$\pm$0.2 Gyr. The good agreement of this result with the 
age of NGC 6362 derived from isochrone comparison with the CMD of the cluster is encouraging. 
This is particularly the case given the age of the cluster as obtained by \citet{dotter10} based 
on an independent photometry (12.5$\pm$0.5 Gyr). However, one should keep in mind that the 
uncertainty of 0.2 Gyr is a purely statistical one, and, as pointed out by \citet{jka13}, it 
should be increased by $\sim$0.85 Gyr to account for the sensitivity of stellar age to uncertainties 
in the chemical composition.

The situation with V41 is less clear. Isochrone fits to $(M-R)$ and $(M-L)$ diagrams of this 
system suggest that it is significantly (by $1.3\pm0.4$ Gyr) younger than V40. 
However, given the recent discovery of two stellar
populations in NGC 6362, such a possibility does not seem to be entirely implausible. Upon 
examining CMDs obtained from HST photometry, \citet{dal14} found that the red giant branch 
of the cluster splits into two sub-branches separated by about 0.1 mag in $(U-V)$, while such effect
is not observed in $(V-I)$. Their Figure 1 explains the split by a difference in light-element 
content. While they use isochrones of the same age, which is reasonable given the relatively 
weak dependence of RGB location on time, they clearly identify the Na-poor populations with the first 
generation of stars, and the Na-rich population with the second one "formed during the first few
$\sim$100 Myr of the cluster life from an intra-cluster medium polluted by the first generation". 
We observed similar effects (i.e. a large split in $(U-V)$ and a much smaller one in $(V-I)$)
using same-age DSED isochrones differing by $\sim$0.1 dex in [Fe/H], or $\sim$0.3 dex in [$\alpha$/Fe], 
or $\sim$0.1 in Y. The RGB also splits when the populations have the same composition but differ in 
age by 4-5 Gyr, however an age difference this large causes the subgiant branch to split even more 
widely than the RGB, while the opposite is observed in NGC 6362. Thus, we corroborate the results 
of \citet{dal14} and their scenario involving two star formation episodes, with the second-generation 
stars slightly enriched in metals (and possibly in helium). 
 
The discrepancy of the ages of V41 components ($1.5\pm0.7$ Gyr) may be regarded as statistically 
insignificant. However we decided to study the possible origin of the difference
by varying chemical composition 
parameters Y, [Fe/H] and [$\alpha$/Fe] (only one of them was varied at a time), and looking for
isochrones joining locations of the primary and the secondary in the $(M-R)$ and $(M-L)$ diagrams. 
We found acceptable (albeit only marginally) fits with Y = 0.26 at an age of 9.5 Gyr; [$\alpha$/Fe] = 0.2 
at an age of 9.7 Gyr, and [Fe/H] = -0.90 at an age of 11.3 Gyr. 
This suggests that V41 may have formed from material 
enriched in products of stellar nucleosynthesis, and thus it may indeed be younger than V40. We 
note parenthetically that the results of our fitting experiments are in line with the findings 
of \citet{jka13}, who, while discussing $(M-R)$ diagrams for binaries in M4, wrote ``increasing 
Y \emph{decreases} the age derived from the $(M-R)$ plane while increasing [Fe/H] or [$\alpha$/Fe] 
\emph{increases} the age.'' 

V40 is also somewhat peculiar because of the eccentricity of its orbit. 
At an age of $\sim$12 Gyr and a period 
of 5.3 d the orbit should long ago have been fully circularized by  tidal friction 
\citep{maz08,mat04}. There is no evidence for the presence of a third body in this system, 
which means that it must have undergone a close encounter during the last few Gyr. According 
to \citet{dal14} NGC 6362 may have had a complicated dynamical history, losing perhaps as much as 
80\% of its original mass. Since close encounters account for a few per cent of the total mass 
loss (M. Giersz, private communication), the very existence of the nonzero eccentricity of V40 
lends some support to this conjecture. 

While V40 and V41 seem to belong to distinct populations, we feel it is too early to
claim a direct detection of non-coevality among stars in globular clusters.
It is clear, however, that DEBs offer the potential to test theories describing the origin 
of multiple populations in GCs, and to verify estimates of age differences between 
populations based on CMD-fitting. For the latter, both the theoretical predictions and 
fitting procedures yield values ranging from $\sim10^7$~yr to several hundred Myr \citep[]
[and references therein]{nar15}, and any independent constraints are highly desirable. 
The most obvious test for the reality of the age difference between our two systems would be to 
determine the chemical composition of their components using disentangling software 
\citep[see e.g.][]{had09}. The existing spectra do not have adequate quality to measure abundances, 
but, given the brightnesses of the components and the orbital periods, it should 
be possible to improve the S/N ratio.

There is also room for improving the accuracy of parameter measurements.
As noted by \citet{jka13}, the errors in the masses of the components originate almost entirely 
from the orbital solution, which may only be improved by taking additional spectra (preferably 
with the same instrument). Admittedly, this would require a substantial observational effort, as doubling 
of the present set of radial velocity data would lead to an improvement of only 33\% in 
the mass estimates \citep{tho10}.
The errors in the radii of the components are dominated in turn by the 
photometric solution, whose quality depends on the accuracy of the photometry.  
We estimate that an accuracy of 0.002 -- 0.003 mag in $V$ would reduce the errors of 
the radii by 50\%. This goal could be easily achieved on a 6-8~m class telescope equipped with 
a camera capable of good PSF sampling. 
Finally, IR photometry should be obtained to reduce errors in luminosity with the help  
of  existing accurate calibrations linking $(V-K)$ color to $V$-band surface brightness.

\acknowledgments
This series of papers is dedicated to the memory of Bohdan Paczy\'nski.
JK and MR were supported by the grant DEC-2012/05/B/ST9/03931 from the 
Polish National Science Center. AD received support from the Australian 
Research Council under grant FL110100012.

\section*{Appendix: Kwalex: A method for timing eclipses}
\citet{kwee} proposed an analyis of the timing of eclipses by fitting them with
symmetric curves, so that any asymmetries would be reflected in observed-minus-calculated 
$(O-C)$ deviations.
Their classical method does have its drawbacks. As far as the algorithm is concerned:
(i) the interpolation of a light curve
into a somewhat arbitrary evenly spaced time mesh yields a slight dependence of
a final estimate of the central time on its initial approximation and (ii)
the resulting estimate of the eclipse centre depends on the initial selection of data points and
on the initial guess of the eclipse centre. There are also statistical drawbacks:
(iii) interpolation effectively assigns uneven weights to points and (iv)
the effective model, a polyline, suffers from excess number of parameters
describing all nodes. Such a waste of degrees of freedom is known in statistics to decrease accuracy.

Let us adopt the zero point of the time scale as the current estimate of the eclipse
centre $T_c^{(n)}$. Following \citet{kwee} we fit
both the input light curve and its reflexion with respect to $T_c^{(n)}$.
The novelty here is that we employ the orthogonal polynomial approximation of the
combined light curve (e.g. \citealt{ral65}),rather than on a poly-line
interpolation. In doing so we optimize the
sensitivity of our method by keeping the number of model parameters at a
statistically justifiable minimum (Occam razor). By virtue of the Fisher
lemma (e.g. \citealt{bra75}), parameters of the orthogonal model are
uncorrelated with noise hence they yield the minimum variance model estimate.

Next, again following \citet{kwee}, we freeze the model curve and shift $T_c^{(n)}$
by $\pm\delta$ and calculate $\chi^2$ of $O-C$ for the time shifts
$-\delta,\;0,+\delta$. These 3 values of $\chi^2$ are fitted with a parabola
\begin{equation}
 \chi^2(t)= at^2+bt+c\label{e1.0}
\end{equation}
and the location of its minimum $T_c^{(n+1)}=-b/2a$ may be adopted as the next approximation of $T_c^{(n+1)}$.
However, to assure convergence, we trim time steps $dT_c=T_c^{(n+1)}-T_c^{(n)}$ and
carefully select  $\delta$ to avoid going too far away from the
solution, yet possibly ensuring bracketing it within $\pm\delta$. Such orthogonal polynomials
retain only even powers and are optimal both in terms of degrees of freedom and  
numerical procedure relying on orthogonal projection. By the
incorporation of weight coefficients in the scalar product our algorithm is able
to account for uneven errors of data.

Complications, both conceptual and algorithmic, arise due to our ultimate aim
of minimizing the dependence of the final result on somewhat accidental
selection of the initial data points. Discrete selection/modification of observations may result in
numerical instability and/or non-converging limit cycles. To minimize
such danger we resort to smooth down-weighting of
points near edge of eclipse by a symmetric Fermi-Dirac distribution function
(e.g. \citealt{kit96}):
   \begin{equation}
 f(t) = \frac{1}{e^{\frac{|t|-\Delta}{\delta}}+1}\label{e1.1}
\end{equation}
The final weight of data measured at time $t_i$ with an error $\sigma_i$  is
$w_i=f(t_i)/\sigma_i^2$. As our time zero-point moved during iterations towards
the final solution, so our weights  vary between iterations. We roughly
followed principles of the minimization algorithm by simulated annealing
\citep{ott89}. At the start of iterations we adopt a rather crude model curve for
$m=2$ and smooth weighting $\delta=\Delta/10$. Note, that in statistical
physics $\delta$ would be proportional to the temperature.  Only after
successful progress of our algorithm do we tighten our data sample by decreasing
$\delta$ (i.e. 'lower temperature') and refine our model by increasing $m$.
Also, we trimmed the time step $dT_c\leq\Delta/10$.

After each iteration we calculate two $T_c$ error estimates: $\Sigma$ and
$\sigma$. The error $\Sigma$ corresponds to such a model shift $\delta=\Sigma$
that $\chi^2$ grows over its minimum value $\chi^2_{min}$ by its variance
$\sqrt{2d}$, where $d=n-m/2-1$ is the number of degrees of
freedom. The error $\sigma$ is defined similarly, for growth by $1$. For this purpose $\chi^2$ has to be
renormalized by a factor $C$ such, that $E\{C\chi^2\}=d$. For an approximate normalization we substitute
the minimum value of the parabola in Eq. (\ref{e1.0}) for its expectation, so that
$C=(4ac-b^2)/4ad$. $\Sigma$ and $\sigma$ are calculated by solving of
Eq. (\ref{e1.0}). In this way we obtain: $a\Sigma^2= \sqrt{2d}\chi^2_{min}/d$ and similarly for $\sigma^2$:
   \begin{eqnarray}
\Sigma^2&=&\sqrt{2}\frac{4ac-b^2}{4a^2\sqrt{d}}\label{e1.2}\\
\sigma^2&=&\frac{4ac-b^2}{4a^2d}\label{e1.3}
\end{eqnarray}
To force a positive result we apply Eq. (\ref{e1.0}) to $\log{\chi^2}$ rather then to $\chi^2$.
In statistical terms, $\Sigma$ corresponds to the maximum reasonable uncertainty of $T_c$, were other parameters strongly
correlated with it and $\sigma$ is its likely error, assuming weak correlation.

During iterations we keep $\delta$ as
large as $\delta=2\Sigma$.  Thus $\delta$ decreases for an improved model fit at
a moderate rate. The polynomial order is raised in steps of $2$, up to its
maximum allowed value, only after decrease of corrections of $T_c$ to
$dT_c\equiv T_c^{(n+1)}-T_c^{(n)} < 0.01\Sigma$. We terminate iterations when
$m$ reaches its selected maximum value and time steps become small $dT_c < \epsilon\Delta$, where
$\epsilon$ is our tolerance parameter. Our result is the final eclipse centre $T_c$ and
its likely error $\sigma$. Note, that for correlated observations errors need to be increased \citep{asc91}.

\clearpage

\begin{deluxetable}{lccc}
\tablecolumns{4}
\tablewidth{0pt}
\tabletypesize{\normalsize}
\tablecaption{Times of minima for V40 
   \label{V40_min}}
\tablehead{
\colhead{E}   &
\colhead{HJD-2450000}     &
\colhead{$\sigma$}    &
\colhead{$O-C$}  
}
\startdata

       -480.5  &     1349.66400 &     0.00129 &    0.00060 \\
       -274.5  &     2440.67681 &     0.00083 &   -0.00019 \\
          0.5  &     3891.82809 &     0.00027 &    0.00045 \\
          3.5  &     3907.71733 &     0.00024 &   -0.00027 \\
         16.5  &     3976.56802 &     0.00139 &   -0.00069 \\
         77.5  &     4299.63424 &     0.00026 &   -0.00024 \\
        138.5  &     4622.70052 &     0.00025 &    0.00015 \\
        276.5  &     5353.57276 &     0.00031 &    0.00003 \\
           0.  &     3915.50987 &     0.00028 &   -0.00022 \\
           1.  &     3920.80387 &     0.00096 &    0.00196 \\
         139.0 &     4651.67750 &     0.00035 &    0.00014 \\
         335.0 &     5689.72753 &     0.00039 &    -0.00007 \\
\enddata
\end{deluxetable}

\clearpage


\begin{deluxetable}{lccc}
\tablecolumns{4}
\tablewidth{0pt}
\tabletypesize{\small}
\tablecaption{Velocity observations of V40
   \label{V40vel}}
\tablehead{
\colhead{HJD-2450000}   &
\colhead{$v_{p}$ [km s$^{-1}]$}    &
\colhead{$v_{s}$ [km s$^{-1}]$}     &
\colhead{phase}
}
\startdata
2870.5433 &   52.73 &  -78.06 &  0.665 \\
3183.6775 &   57.18 &  -85.40 &  0.790 \\
3201.5720 &  -74.19 &   52.16 &  0.169 \\
3206.5845 &  -65.72 &   41.55 &  0.115 \\
3210.5848 &   35.22 &  -64.64 &  0.871 \\
3517.7456 &   36.05 &  -63.77 &  0.867 \\
3521.7538 &   40.02 &  -66.70 &  0.624 \\
3580.5437 &   60.61 &  -87.52 &  0.725 \\
3585.6468 &   56.59 &  -84.07 &  0.688 \\
3586.6453 &   33.28 &  -59.01 &  0.877 \\
3815.8955 &  -74.88 &   52.41 &  0.163 \\
3816.8897 &  -62.77 &   41.06 &  0.350 \\
3877.8517 &   38.86 &  -66.45 &  0.861 \\
3890.7002 &  -74.76 &   55.72 &  0.287 \\
3935.6890 &   57.56 &  -86.79 &  0.782 \\
3937.6503 &  -72.90 &   50.94 &  0.152 \\
3938.5635 &  -68.71 &   47.47 &  0.324 \\
3991.5267 &  -68.84 &   47.33 &  0.325 \\
4258.7016 &   59.64 &  -88.13 &  0.771 \\
4258.7465 &   57.69 &  -86.47 &  0.780 \\
4258.7904 &   58.16 &  -85.71 &  0.788 \\
4329.5454 &  -72.97 &   49.67 &  0.148 \\
4648.6241 &  -49.91 &   27.55 &  0.395 \\
4656.6819 &   15.95 &  -41.39 &  0.916 \\
\enddata
\end{deluxetable}

\clearpage


\begin{deluxetable}{lccc}
\tablecolumns{4}
\tablewidth{0pt}
\tabletypesize{\small}
\tablecaption{Velocity observations of V41
   \label{V41vel}}
\tablehead{
\colhead{HJD-2450000}   &
\colhead{$v_{p}$ [km s$^{-1}]$}    &
\colhead{$v_{s}$ [km s$^{-1}]$}     &
\colhead{phase}
}
\startdata
2871.5571 &  -32.73 &    8.93 &  0.181 \\
2872.5519 &  -38.27 &   15.99 &  0.236 \\
3179.7609 &  -47.66 &   27.81 &  0.410 \\
3206.5336 &   20.39 &  -52.18 &  0.906 \\
3517.7940 &  -43.64 &   23.06 &  0.306 \\
3520.7452 &  -46.33 &   26.13 &  0.471 \\
3521.7074 &  -40.89 &   20.15 &  0.525 \\
3580.6885 &   42.56 &  -73.79 &  0.822 \\
3581.6731 &   30.56 &  -59.28 &  0.877 \\
3582.5946 &   16.30 &  -44.91 &  0.928 \\
3891.6858 &  -34.69 &   14.52 &  0.207 \\
3937.6050 &   44.58 &  -76.40 &  0.774 \\
3938.6590 &   40.59 &  -73.04 &  0.833 \\
3990.5131 &   35.50 &  -66.03 &  0.731 \\
4329.6373 &   15.02 &  -42.95 &  0.689 \\
4648.7184 &  -41.08 &   19.93 &  0.525 \\
4649.6742 &  -30.79 &    8.04 &  0.579 \\
4649.7192 &  -30.27 &    7.02 &  0.581 \\
4652.7191 &   40.41 &  -73.29 &  0.749 \\
4655.7155 &   19.30 &  -48.17 &  0.917 \\
4967.7970 &  -47.13 &   25.65 &  0.362 \\
5012.6510 &   30.80 &  -62.41 &  0.870 \\
5038.5921 &  -44.61 &   23.99 &  0.320 \\
6842.6401 &  -30.67 &    7.75 &  0.167 \\
6844.6397 &  -41.48 &   20.30 &  0.279 \\
6845.6556 &  -46.30 &   24.06 &  0.336 \\
\enddata
\end{deluxetable}

\clearpage
\nopagebreak


\begin{deluxetable}{lcc}
\tablecolumns{3}
\tablewidth{0pt}
\tabletypesize{\normalsize}
\tablecaption{Orbital and photometric parameters\tablenotemark{a} of the systems V40 and V41
   \label{param}}
\tablehead{
\colhead{Parameter}   &
\colhead{V40}     &
\colhead{V41}   
}
\startdata
     $\gamma$ (km s$^{-1}$)     & -12.34(11)       & -12.55(8)  \\
     $K_p$ (km s$^{-1}$)        & 70.18(17)        &  46.59(14) \\
     $K_s$ (km s$^{-1}$)        & 73.63(25)        &  52.57(17) \\
     $\sigma_p$ (km s$^{-1}$)   & 0.72             &   0.53    \\
     $\sigma_s$ (km s$^{-1}$)   & 1.07             &   0.65    \\
     $A$   (R$_\odot$)          &  15.037(32)      &   33.294(71) \\
     $e$                        & 0.05054(73)      & 0.3125(13)  \\
     $\omega$ (deg)             & 206.6(17)        & 321.23(33)   \\
     $i$ (deg)     & 88.22(7)   & 89.547(20)    \\
     $r_p$         & 0.08817(45)  & 0.03225(13)   \\
     $r_s$         & 0.06627(87)  & 0.02195(12)   \\
     $S_V$         & 0.9815(35)     &  0.7333(63)  \\
     $S_B$         & 0.9686(18)      &  0.669(11)     \\
     $(L_p/L_s)_V$       & 1.803(54)   & 2.964(27)   \\
     $(L_p/L_s)_B$       & 1.8272(57 ) & 3.264(44)   \\
     $\sigma_\mathrm{rms}(V)$ (mmag)
                   & 11             &   12        \\
     $\sigma_\mathrm{rms}(B)$ (mmag)
                   & 13            &    16        \\
     $V_p$ (mag)\tablenotemark{b}   & 18.698(13)(20) & 19.089(7)(17)  \\
     $V_s$ (mag)\tablenotemark{b}   & 19.338(22)(27) & 20.274(10)(18)  \\
     $(B-V)_p$ (mag)\tablenotemark{b}   & 0.542(18)(23) & 0.550(10)(18)  \\
     $(B-V)_s$ (mag)\tablenotemark{b}   & 0.556(31)(34) & 0.650(16)(22) \\
\enddata
\tablenotetext{a}{Numbers in parentheses are the errors of the 
last significant digit(s).}
\tablenotetext{b}{For $V$ and $(B-V)$ both the 
internal error (from the photometric solution and profile photometry) 
and the total error is given,
the latter including 0.015 mag uncertainty of the zero point of the magnitude 
              scale.}
\end{deluxetable}

\clearpage


\begin{deluxetable}{lcc}
\tablecolumns{4}
\tablewidth{0pt}
\tabletypesize{\normalsize}
\tablecaption{Absolute parameters\tablenotemark{a} of the components 
 of V40 and V41
   \label{param2}}
\tablehead{
\colhead{Parameter}   &
\colhead{V40}     &
\colhead{V41}    
}
\startdata
     $M_p$ (M$_\odot$)  & 0.8337(63) & 0.8215(58)  \\
     $M_s$ (M$_\odot$)  & 0.7947(48) & 0.7280(47)  \\
     $R_p$ (R$_\odot$)  & 1.3253(77) & 1.0739(48) \\
     $R_s$ (R$_\odot$)  & 0.997(13) & 0.7307(46)  \\

     $T_p$ (K)          & 6156(123) & 6124(121)  \\
     $T_s$ (K)          & 6100(158) & 5747(122) \\ 
     $L^\mathrm{bol}_p$ 
           (L$_\odot$)  &2.27(18)  &1.46(11) \\
     $L^\mathrm{bol}_s$ 
           (L$_\odot$)  &1.24(13)  &0.524(45)\\
     $M_{V_p}$ 
           (mag)        &3.92(8)  & 4.40(8)   \\
     $M_{V_s}$ 
           (mag)        &4.59(11) & 5.54(9)    \\
     $(m-M)_{V_p}$ 
           (mag)        & 14.78(8)& 14.69(8)   \\
     $(m-M)_{V_s}$ 
           (mag)        & 14.75(9) & 14.73(8)  \\
     $Age_p$ (Gyr)      & 11.7(3)  &10.7(3)    \\
     $Age_s$ (Gyr)      & 11.5(4)  &9.2(6)     \\
     $Age_b$ (Gyr) &11.7(2) &10.4(3)    \\
\enddata
\tablenotetext{a}{Numbers in parentheses are the errors of the last 
significant digit(s). The age of the binary $Age_b$ is a weighted mean 
of component ages.}

\end{deluxetable}

\clearpage


\clearpage


\clearpage


\clearpage


\clearpage


\clearpage


\clearpage


\end{document}